\documentclass[12pt]{article}
\usepackage{graphicx,graphics,pifont,amssymb}
\usepackage[figuresright]{rotating}

\begin{document}

\pagestyle{empty}

\begin{titlepage}

\title{Flow visualization of an emulsion in a tilted rotating tank
\author{David Swan$^1$ \& Andrew White$^2$ \& Dr. Thomas Ward$^2$\\
\small{\emph{$^1$Department of Nuclear Engineering, North Carolina State University}} \\
\small{\emph{Raleigh, NC 27695-7909}} \\
\small{\emph{$^2$Department of Mechanical and Aerospace Engineering, North Carolina State University}} \\
\small{\emph{Raleigh, NC 27695-7910}}}}
\date{} \maketitle

\begin{abstract}

Inhomogeneous fluid mixing in a tilted-rotating cylindrical tank (radius $a=3.5$ cm) is shown at ~Re(17-40) and low capillary numbers. A water and surfactant solution (1\% by mass sodium oleate) is dispersed in soybean oil (95\% by volume), through varying the rotation rate, $\Omega$ and angle of inclination, $\theta$ the rate of mixing is observed. A planar laser is directed down the tank axis to highlight a cross-sectional area of the fluid volume and as the water droplets begin to break up to sizes on the order of the beam width and less, more light is refracted and the mixture is illuminated. Initially, the water breaks up into large droplets that exhibit approximate solid-body rotation about the bottom of the tank. When the total combined volume is below the critical volume of the tank $V_{crit} = \pi a^3 \tan \theta$ vortex transport of the water occurs more rapidly, breaking up the water into continually smaller droplets in a process that resembles periodic shearing. When the fluid volume is above critical the water will break up and rotate about the bottom of the tank and vortex-induced mixing is much more reticent, if occurring at all. It is noted that shallower angles with respect to the horizontal produce faster mixing while allowing a greater volume of fluid to be mixed at the sub-critical volume given a constant tank size.\\

This method of mixing bears relevance to increasing efficiency in creating diesel-fuel-and-water emulsions. Suspending water in diesel has been shown to increase combustion efficiency and moderate pollutant production to a significant degree [Crookes et al., Energy Convers. Mgmt., 1997] while also allowing for the addition of water-soluble fuel additives. The advent of the tilted-rotating tank offers a decrease in complexity and power consumption during this process. Also of note is a potential for employ in the nuclear industry with respect to spent-fuel reprocessing; the Taylor-Couette device currently used to perform aqueous separation of fuel material could perhaps be optimized with the use of a tilted rotating tank.

\end{abstract}

\noindent E-mail: tward@ncsu.edu

\end{titlepage}

\end{document}